\documentclass[a4paper,11pt]{article}
\pdfoutput=1 % if your are submitting a pdflatex (i.e. if you have
\usepackage{jheppub} % for details on the use of the package, please
\usepackage[T1]{fontenc} % if needed

\makeatletter
\gdef\@fpheader{}
\makeatother
\vspace*{0.1cm}

\title{\boldmath Filtered asymmetric dark matter during the Peccei-Quinn phase transition}

\author{M. Ahmadvand}

\affiliation{School of Particles and Accelerators, Institute Research in Fundamental Sciences (IPM), P. O. Box 19395-5531, Tehran, Iran}
\emailAdd{ahmadvand@ipm.ir}
\abstract{In this paper, we propose a bubble \textit{filtering-out} mechanism for an asymmetric dark matter scenario during the Peccei-Quinn (PQ) phase transition. Based on a QCD axion model, extended by extra chiral neutrinos, we show that the PQ phase transition can be first order in the parameter space of the model and regarding the PQ symmetry breaking scale, the mechanism can generate PeV-scale heavy neutrinos as a dark matter candidate. Considering a CP-violating source, during the phase transition, discriminating between the neutrino and antineutrino number density, we find the observed dark matter relic abundance, such that the setup can be applied to the first order phase transition with different strengths. We then calculate effective couplings of the QCD axion addressing the strong CP problem within the model. We also study the energy density spectrum of gravitational waves generated from the first order phase transition and show that the signals can be detected by future ground-based detectors such as Einstein Telescope. In particular, for a visible heavy axion case of the model, it is shown that gravitational waves can be probed by DECIGO and BBO interferometers. Furthermore, we discuss the dark matter-standard model neutrino annihilation process as a source for the creation of PeV-scale neutrinos.}


\begin{document} 
\maketitle
\flushbottom

\section{Introduction}
\label{intro}

Astrophysical and cosmological observations including measurements of galactic rotation curves, gravitational lensings, and anisotropies in the Cosmic Microwave Background (CMB) support the existence of Dark Matter (DM) forming around 26\% of the Universe energy density \cite{Bertone:2016nfn}. However, the nature of DM in fundamental physics is still unknown.

To explain the relic abundance of the cold DM from the early Universe, various mechanisms and candidates have been suggested. Due to the coincidence between the weak scale and massive particles with a mass range from GeV to TeV, and also relying on the cosmological expansion and thermal freeze-out mechanism, Weakly Interacting Massive Particle (WIMP) paradigm has been an attractive scenario \cite{Bertone:2004pz}, though direct detection searches disfavor some of these scenarios and lead us to consider other possibilities such as non-thermal DM models \cite{Allahverdi:2010rh} and scenarios in which DM is produced in the decay of thermally-decoupled heavy particles \cite{Ahmadvand:2020izy, Kolb:1998ki}. Another remarkable fact about the DM energy density is its closeness to the abundance of asymmetric baryonic matter. This may imply a common asymmetric origin for dark and visible matter and is the motivation for asymmetric DM scenarios \cite{Ahmadvand:2020izy, Kaplan:2009ag}.

In this paper, we use the so-called filtered DM mechanism through which DM dynamically acquires a mass and its number density is abruptly frozen out \cite{Baker:2019ndr, Chway:2019kft}. Based on first order Phase Transitions (PTs) and DM interactions with bubbles, this mechanism provides a framework to evade the Griest-Kamionkowski (GK) bound on the mass of thermally-produced DM and a possibility to produce DM masses above $100\, \mathrm{TeV}$.

On the other hand, at these high energy scales, one of the well-motivated cosmological PTs could have taken place. In the early Universe at temperatures around $100\,\mathrm{PeV}$, the Peccei-Quinn (PQ) symmetry $\mathrm{U(1)}_{\mathrm{PQ}}$ is spontaneously broken according to QCD axion models, resolving the strong CP problem \cite{Peccei:1977hh,Peccei:2006as}.\footnote{Considering astrophysical constraints, the symmetry breaking scale is bounded between few $ 10^8\,\mathrm{GeV}$ and few $ 10^{17}\,\mathrm{GeV}$ \cite{Raffelt:2006cw,Arvanitaki:2009fg}.} Provided that the PQ PT is first order, the highly massive DM candidates can be naturally generated through the bubble filtering-out mechanism. Moreover, in this work according to this setup, we propose an asymmetric DM scenario during the PQ PT.

We use a Dine-Fischler-Srednicki-Zhitnitsky (DFSZ) axion model \cite{Dine:1981rt,Zhitnitsky:1980tq} supplemented with chiral neutrinos, where one of the flavors can play the role of DM and we focus on this flavor in our discussions. Taking loop quantum effects into account at finite temperature, within the parameter space of the theory we show that the PQ PT can be first order and consequently bubbles of the broken phase can nucleate and expand. DMs whose kinetic energy is greater than their masses in the broken phase enter the bubbles. In addition, Right-Handed (RH) neutrino interactions with bubbles violate the lepton number. We also consider the possibility of a CP violation source varying during the PT \cite{Baldes:2016gaf,Cline:2020jre,Long:2017rdo}. Due to CP-violating effects, we obtain the net dark number density and hence the observed relic abundance can be found and survive after the PT. Moreover, the baryon asymmetry can be fulfilled due to asymmetric visible neutrinos via leptogenesis scenarios \cite{Ahmadvand:2020izy,Davidson:2008bu}.\footnote{For baryogenesis scenarios fulfilled via lepton-number violation at a first order PT, see \cite{Long:2017rdo,Pascoli:2016gkf,Cohen:1990py,Cohen:1990it}.}

We then explore phenomenological consequences of the model. As a solution to the strong CP problem, we study effective interactions of the QCD axion in the model. Regarding the generation of Gravitational Waves (GWs) from first order PTs, GW direct detection experiments are powerful tools to probe early Universe events. By obtaining the bubble profile and other required quantities, including the vacuum energy and duration of the PT, we find the energy density spectrum of GWs produced from the first order PQ PT. Furthermore, we discuss the detectability of possible GWs from the PT in the case of visible QCD axions  with PQ symmetry breaking scale around $100\,\mathrm{TeV}$ \cite{Rubakov:1997vp,Berezhiani:2000gh,Hook:2014cda,Fukuda:2015ana}. Finally, we investigate some of effective DM interactions with Standard Model (SM) particles.

In Section \ref{model}, we introduce the model and discuss the PQ PT. In section \ref{relic}, the DM relic abundance is obtained within the model. We study phenomenological aspects of the model in Section \ref{phen} and conclude in Section \ref{con}.

\section{The Model}
\label{model}
Due to the anomalous U(1) axial symmetry of strong interactions and QCD vacuum structure, an effective CP-violating term associated with $\theta$-vacuum is allowed in the Lagrangian. This term contributes to the neutron electric dipole moment, $d_n\approx3.6\times 10^{-16}\bar{\theta}\,e\, \mathrm{cm}$ \cite{Crewther:1979pi}, which is experimentally constrained $|d_n|<2.9\times 10^{-26}\,e\, \mathrm{cm}~(90\%~\mathrm{C.L.})$ \cite{Baker:2006ts} and thereby $\bar{\theta}\lesssim 10^{-10} $. This poses the strong CP problem in that there is no reason in the SM why $ \bar{\theta}$ should be very small. 

An interesting solution to this problem is based on a global chiral $\mathrm{U(1)}_{\mathrm{PQ}}$ symmetry proposed by Peccei and Quinn \cite{Peccei:1977hh}. In fact, at QCD scales, the interaction of a pseudo-scalar field, the axion $a$, is effectively added to the Lagrangian, $ (a/f_a +\bar{\theta})G\widetilde{G} $, where $f_a$ is the axion decay constant, $G_{\mu\nu}$ is the gluon field strength and $\widetilde{G} $ denotes its dual. Therefore, the CP-violating $\bar{\theta}$-term can be cancelled via the vacuum expectation value (vev) of axion at the minimum of its potential, addressing the strong CP problem. The UV completion of such non-renormalizable interactions can be constructed in a model invariant under $\mathrm{U(1)}_{\mathrm{PQ}}$. At some high energy scale, the symmetry is spontaneously broken and the resulting pseudo-Nambu Goldstone boson would be matched with the axion so that at low energies the effective interaction can be induced due to the chiral anomaly and QCD instantons. 

In general, considering astrophysical bounds on the PQ symmetry breaking scale, two types of axion models can be categorized: Kim-Shifman-Vainshtein-Zakharov (KSVZ) models \cite{Kim:1979if,Shifman:1979if} which contain extra heavy quarks and PQ scalar fields, carrying the PQ charge, and DFSZ models \cite{Dine:1981rt,Zhitnitsky:1980tq} in which beside the PQ scalar field, an additional Higgs field is introduced. We here employ a DFSZ type axion model extended by three RH neutrinos, where one of the flavors can be regarded as a DM candidate.

The Lagrangian of the model is given by
\begin{equation}\label{la}
	\mathcal{L}\supset |\partial_{\mu}\Phi|^2+V(\Phi,H_u,H_d)+\overline{N}_{R_i}i\partial_{\mu} N_{R_i}+y_{ij}\,\Phi\overline{N}_{R_i}N_{R_j}^c
\end{equation}
where  

\begin{equation}
	\begin{aligned}
		V(\Phi,H_u,H_d) &=\lambda_{\phi}\left(|\Phi|^{2}-v_{\phi}^{2} / 2\right)^{2}+\left|H_{d}\right|^{2}\left(\kappa_{d}|\Phi|^{2}-\mu_{d}^{2}\right)+\left|H_{u}\right|^{2}\left(\kappa_{u}|\Phi|^{2}+\mu_{u}^{2}\right) \\
		&-\left(\kappa \Phi^{\dagger} H_{u} H_{d}+h . c .\right)+\lambda_{d}\left|H_{d}\right|^{4}+\lambda_{u}\left|H_{u}\right|^{4}+\lambda_{1}\left|H_{u} H_{d}\right|^{2}+\lambda_{2}\left|H_{u}\right|^{2}\left|H_{d}\right|^{2}
	\end{aligned}
\end{equation}
 $ \Phi $ is the PQ scalar field singlet under the SM gauge symmetry group $\mathrm{SU(2)}_L\times\mathrm{U(1)}_Y$, $ H_u$ and $H_d$ denote two $\mathrm{SU(2)}_L$ doublets, and the interaction of singlet RH neutrinos with $ \Phi $ violates the lepton number by two units $ \Delta L=2$ in case $L(N_{R_i})=1$.\footnote{We assume the coupling of such interactions $\Phi\overline{N}_LN_R $ is very small compared to that of L-violating Yukawa interactions.} We also consider a discrete $ \mathbb{Z}_2$ symmetry under which all fields are even except for the dark neutrino flavor which is odd so that the massive DM gets thermally-decoupled after the PT. We assume that $H_u$ interacts with $u$ quarks and the rest of the SM as well as two visible RH neutrinos are coupled to $H_d$
\begin{equation}
y_u\overline{Q}_L H_u u_R+y_d\overline{Q}_L H_d d_R+y_l\overline{\Psi}_{L_l} H_d l_R+y_{N_{\alpha}}\overline{\Psi}_{L_l} \widetilde{H}_d N_{R_{\alpha}}+h.c.. 
\end{equation}
The Lagrangian should be invariant under $\mathrm{U(1)}_{\mathrm{PQ}}$. PQ invariance of $\Phi^{\dagger} H_u H_d $ implies $X_{H_u}+X_{H_d}=X_{\Phi} $ where $X_X$ is the PQ charge of a given field. Moreover, imposing the orthogonality between PQ and corresponding hypercharge currents, $ -X_{H_u}v_u^2+X_{H_d}v_d^2=0 $ \cite{DiLuzio:2020wdo}. Defining $v_u=v\sin\theta $ and $v_d=v\cos\theta $, where $v$ is the Electroweak (EW) vev, and choosing $X_{\Phi}=1 $, we may determine all charges through
\begin{eqnarray}
X_{H_d}=\sin^2\theta,~~~~~X_{H_u}=\cos^2\theta,
\end{eqnarray} 
as shown in Table \ref{t1}. The SM Higgs is indeed $ H=H_d\cos\theta+\widetilde{H}_u\sin\theta $, where $ \widetilde{H}_u=i\sigma_2H^*_u $, and $ v^2=v_d^2+v_u^2$. Given that $v_a^2=\sum_i X_iv_i^2=v_{\phi}^2+v^2\sin^2 (2\theta)/4 $ and $v_{\phi}\gg v $, we have $ v_a\simeq v_{\phi}$. We will also show in Section \ref{phen} that $ v_a=f_a$ in the model. The PQ symmetry is spontaneously broken by the vev of $ \Phi$, such that $ \Phi=1/\sqrt{2}(\phi+f_a)\exp(i a/f_a)$ and the axion field is shifted by the PQ transformation.
\begin{table}
	\begin{center}
		\begin{tabular}{|c|c|c|c|c|}
			\hline $\text { Field }$ & $\mathrm{SU(3)}_{c}$ & $\mathrm{SU(2)}_{L}$ &$ \mathrm{U(1)}_{Y}$ & $\mathrm{U(1)}_{\mathrm{PQ}}$ \\
			\hline $\Phi$ & $\mathbf{1}$ & $\mathbf{1}$ & 0 & 1 \\
			\hline 	$H_{u}$ & $\mathbf{1}$ & $\mathbf{2}$ & $-\frac{1}{2}$ & $\cos^2\theta$ \\
			\hline $H_{d}$ & $\mathbf{1}$ & $\mathbf{2}$ & $\frac{1}{2}$ & $\sin^2\theta$ \\
			\hline $Q_{L}$ & $\mathbf{3}$ & $\mathbf{2}$ & $\frac{1}{6}$ & $\cos^2\theta$ \\
			\hline $u_{R}$ &  $\mathbf{3}$ & $\mathbf{1}$ & $\frac{2}{3}$ & 0 \\
			\hline $d_{R}$ & $\mathbf{3}$ & $\mathbf{1}$ & $-\frac{1}{3}$ & 2$\cos^2\theta$-1 \\
			\hline $\Psi_{L}$ & $\mathbf{1}$ & $\mathbf{2}$ & $-\frac{1}{2}$ & $ \frac{1}{2}-\sin^2\theta$ \\
			\hline $e_{R}$  & $\mathbf{1}$ & $\mathbf{1}$ & -1 & $\frac{1}{2}-2\sin^2\theta $ \\
			\hline $ N_{R}$  & $\mathbf{1}$ & $\mathbf{1}$ & 0 & $ \frac{1}{2}$ \\
			\hline
		\end{tabular} 
		\caption{Charge assignment and group representation of the fields in the model}\label{t1}
	\end{center}
\end{table}

\subsection{First order Peccei-Quinn phase transition}\label{first}
In order to describe the PT, we first express the potential in terms of electrically neutral components, $ H_u=h_u/\sqrt{2}$ and $ H_d=h_d/\sqrt{2}$ at the tree-level
\begin{equation}
	\begin{aligned}
		V &=\frac{\lambda_{\phi}}{4}\left(\phi^{2}-f_{a}^{2}\right)^{2}+\frac{1}{2} h_{d}^{2}\left(\frac{\kappa_{d}}{2} \phi^{2}-\mu_{d}^{2}\right)+\frac{1}{2} h_{u}^{2}\left(\frac{\kappa_{u}}{2} \phi^{2}+\mu_{u}^{2}\right)-\frac{\kappa}{\sqrt{2}} \phi h_{u} h_{d} \\
		&+\frac{\lambda_{d}}{4} h_{d}^{4}+\frac{\lambda_{u}}{4} h_{u}^{4}+\frac{\lambda_{1}+\lambda_{2}}{4} h_{u}^{2} h_{d}^{2}.
	\end{aligned}
\end{equation}
We assume the case in which $ h_u$ vanishes during the PT, hence we will not consider its dynamics during the PT, though in the loop level its loop effects appear in the potential for $ \phi$ and $ h_d$. Consequently, there are two minima: 1- ($ \phi=f_a,~~h_d=h_u=0$) and 2- ($h_d=\mu_d^2/\lambda_{d},~~\phi=h_u=0 $). For this PT, we study the first direction.\footnote{About the possibility of a supercooling period during the PT in axion models see \cite{VonHarling:2019rgb,DelleRose:2019pgi,Ghoshal:2020vud}.} To have such a minimum, the potential should satisfy the condition $ V(\mathrm{min}1)<V(\mathrm{min}2)$ from which $\lambda_{\phi}>\mu_{d}^4/(f_a^4\lambda_{d}) $. Moreover, we are interested in the case in which $ m_{N_i}>m_{\phi}$ in the broken phase, thus $ y_i>\sqrt{2\lambda_{\phi}}$, where $ y_i$ stands for the Yukawa couplings of heavy neutrinos in the diagonal form.

\begin{figure}[tbp]
	\centering
	\includegraphics[scale=0.8]{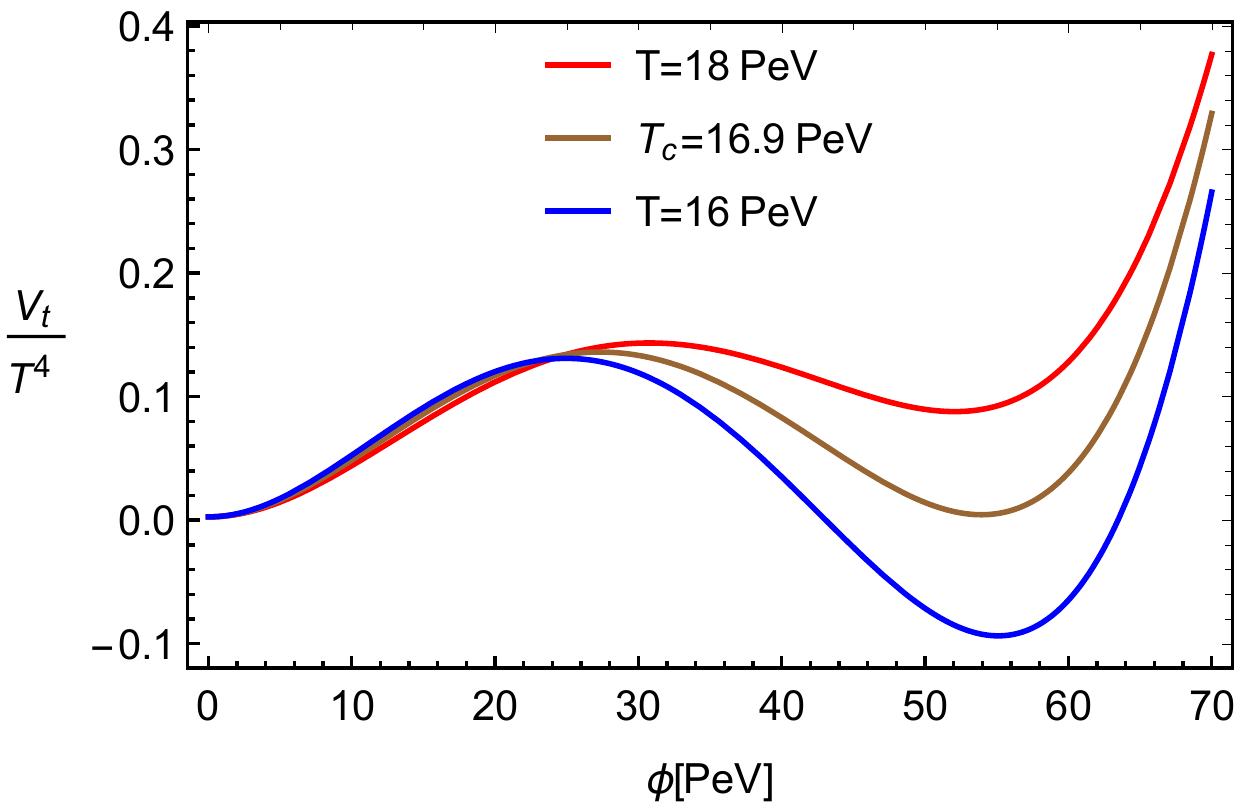}
	\caption{For $ f_a=100\,\mathrm{PeV}$, we display the total potential at three different temperatures within the aforementioned parameter space of the model.}
	\label{f1}
\end{figure}

To account for quantum and thermal effects on the potential, we obtain the one-loop effective potential at finite temperature containing the resummed daisy (ring) diagrams,
\begin{equation}
V_{t}=V+V_{\mathrm{CW}}+V_{\mathrm{th}}+V_{r}
\end{equation}
where at zero temperature, the one-loop Coleman-Weinberg quantum correction can be written as \cite{Coleman:1973jx}
\begin{equation}
V_{\mathrm{CW}}\left(\phi\right)=\sum_{i}(-1)^{F_{b/f}} g_{i} \frac{m_{i}^{4}\left(\phi\right)}{64 \pi^{2}}\left[\ln \left(\frac{m_{i}^{2}\left(\phi\right)}{\Lambda^{2}}\right)-c_{i}\right].
\end{equation}
Here we take $ \Lambda=f_a$, $ F_{b/f}=1(0)$ for fermions (bosons), $ g_i$ is the number of degrees of freedom for a given field, and $c_i=3/2(5/2)$ for scalars and fermions (vectors). Also, the one-loop thermal correction is given by \cite{Quiros:1999jp,Curtin:2016urg}
\begin{equation}
V_{T}\left(\phi, T\right)=\sum_{i}(-1)^{F_{b/f}} g_{i} \frac{T^{4}}{2 \pi^{2}} J_{b / f}\left[\frac{m_{i}^{2}\left(\phi\right)}{T^{2}}\right]
\end{equation}
where 
\begin{equation}
J_{b/f}\left(y^{2}\right)=\int_{0}^{\infty} d x\, x^{2} \ln \left[1 \mp e^{-\sqrt{x^{2}+y^{2}}}\right]
\end{equation}
and the useful form of these thermal functions in the high temperature limit is given by
\begin{equation}\label{therm}
	J_{b}\left(\frac{m^{2}}{T^{2}}\right)=-\frac{\pi^{4}}{45}+\frac{\pi^{2}}{12}\left(\frac{m}{T}\right)^{2}-\frac{\pi}{6}\left(\frac{m^{2}}{T^{2}}\right)^{3 / 2}-\frac{1}{32}\left(\frac{m}{T}\right)^{4} \ln \left(\frac{m^{2}}{a_{b} T^{2}}\right)+\cdots
\end{equation}
\begin{equation}
	J_{f}\left(\frac{m^{2}}{T^{2}}\right)=\frac{7 \pi^{4}}{360}-\frac{\pi^{2}}{24}\left(\frac{m}{T}\right)^{2}-\frac{1}{32}\left(\frac{m}{T}\right)^{4} \ln \left(\frac{m^{2}}{a_{f} T^{2}}\right)+\cdots
\end{equation}
where $ \ln(a_b)=5.4076 $ and $ \ln(a_f)=2.6351 $. The leading order of multi-loop corrections is included in the so-called daisy diagram whose contribution is given by
\begin{equation}
V_{r}\left(\phi, T\right)=-\sum_{i\in b}\frac{g_{i} T}{12 \pi}\left(\left[m_{i}^{2}(\phi)+\Pi_{i}(T)\right]^{3 / 2}-\left[m_{i}^{2}(\phi)\right]^{3 / 2}\right)
\end{equation}
where the thermal mass squared of the scalars are
\begin{equation}
\Pi_{\phi}(T)=\frac{T^2}{6}\left(\kappa_d+\kappa_u+2\lambda_{\phi}\right), 
\end{equation}
\begin{equation}
\Pi_{h_d}(T)=\frac{T^2}{48}\left(9g^2+3g'^2+\frac{12\lambda_t^2}{\cos^2\theta_{\mathrm{w}}}+24\lambda_d+4\kappa_d+8\lambda_{1}+8\lambda_{2}\right), 
\end{equation}
\begin{equation}
\Pi_{h_u}(T)=\frac{T^2}{48}\left(9g^2+3g'^2+24\lambda_u+4\kappa_u+8\lambda_{1}+8\lambda_{2}\right). 
\end{equation}
where $ \lambda_{t}\simeq 1$, $ g\simeq 0.42$, $ g'\simeq 0.12$, and $\cos^2\theta_{\mathrm{w}}\simeq 0.77 $. We calculate the effective potential accordingly with some representative values of parameters, e.g., for $ y_{1}=\sqrt{2}$, $y_{2}=\sqrt{2}/2$, $y_{3}=\sqrt{2}\times 10^{-3}$, $ \kappa_{u}=3$, $ \lambda_{u}=\lambda_{d}=\lambda_{1}=\lambda_{2}=10^{-2}$ and $\kappa_{d}<\lambda_{\phi}=10^{-8} $. As shown in Fig.\ (\ref{f1}), for $f_a=100\,\mathrm{PeV}$, we find the PQ PT is of first order and the two degenerate phases are separated through a barrier at the critical temperature $T_c=16.88\,\mathrm{PeV}$. Based on these parameter values, the first order PT is mainly induced by loop correction terms of $h_u$. Despite the effect of daisy diagram corrections which reduce the altitude of the potential energy barrier and weaken the strength of the PT, the  barrier is generated due to the thermal cubic term, Eq.\ (\ref{therm}).

\begin{figure}[tbp]
	\centering
	\includegraphics[scale=0.9]{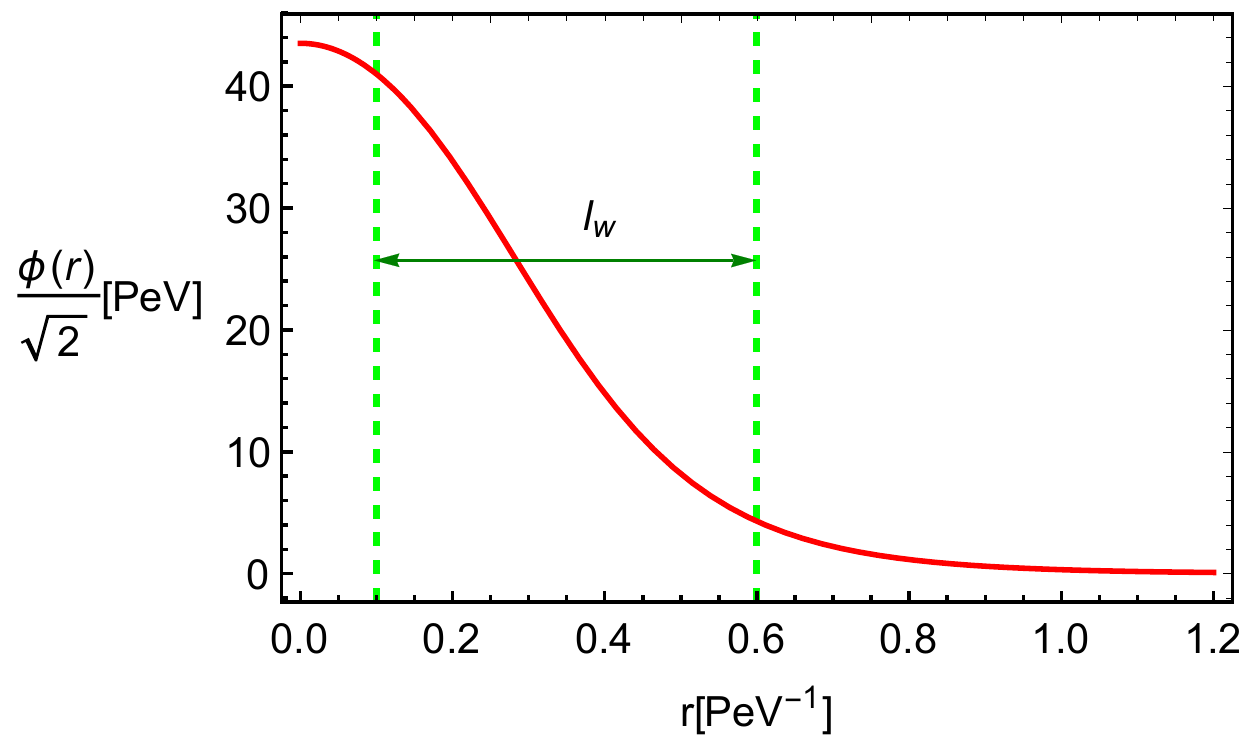}
	\caption{The bubble profile as the solution of the bounce equation is shown. $ l_w$ denotes the wall width.}
	\label{f2}
\end{figure}

At some lower temperature, bubbles of the new phase nucleate, expand and eventually collide with each other until the Universe is completely transmitted into the broken phase. Indeed, bubbles are non-perturbative solutions of the following three-dimensional Euclidean bounce action quantifying the tunneling process
\begin{equation}\label{bac}
S_3(T)=\int_{0}^{\infty}4\pi r^2~dr\Big[\frac{1}{2}\Big(\frac{d\phi}{dr}\Big)^2+V_{t}(\phi, T)\Big]. 
\end{equation}
By extremizing the action, we obtain the bounce equation and according to the following boundary condition, the equation can be numerically solved
\begin{equation}
	\frac{d^2\phi}{dr^2}+\frac{2}{r}\frac{d\phi}{dr}=\frac{\partial V_{t} }{\partial\phi},~~~~~~~\left.\frac{d\phi}{dr}\right|_{r=0}=0,~~~~\phi(\infty)=0.
\end{equation}
Using the \textit{any bubble} code \cite{Masoumi:2016wot}, as can be seen from Fig.\ (\ref{f2}), we obtain the bubble profile connecting the two phases.
Bubbles nucleate when the bubble formation probability, proportional to $ \exp \left(-S_3(T)/T\right)$, is of the order of one. Therefore, we can find the nucleation temperature, $T_n$, by the following relation \cite{Linde:1980tt}
\begin{equation}\label{nuc}
\frac{S_3(T_n)}{T_n}=4\ln\Big(\frac{T_n}{H_*}\Big), 
\end{equation}
where $ H_*\simeq T_n^2/ M_{\mathrm{pl}}  $ is the Hubble parameter and $ M_{\mathrm{pl}}\simeq 2.43\times 10^{18}\,\mathrm{GeV}$ is the reduced Planck mass. As a result, using the obtained solution, from Eq.\ (\ref{nuc}), we find the nucleation temperature of the PT, $ T_n\simeq 9.19\,\mathrm{PeV}$.

In the next section, relying on the first order PQ PT and the obtained parameters, we proceed to calculate the DM relic abundance.

\section{Dark matter relic abundance}
\label{relic}
In this section, using the DM filtration mechanism, we will find the net number density of DM by solving the Boltzmann equation during the PQ PT. According to the first order PT breaking the PQ symmetry, DM acquires a mass $m_{N}^{\mathrm{in}}$ ($ N\equiv N_{\mathrm{DM}}$) and its interactions can be put out of equilibrium inside the bubbles. DM particles and antiparticles that have enough kinetic energy, greater than $m_{N}^{\mathrm{in}}$, enter the bubbles and otherwise reflected DMs remain in equilibrium with the thermal bath. Furthermore, we are interest in the possibility of a varying CP-violating source, similar to the varying bubble profile, during the PT and because of the L-violating interaction of DMs with bubbles, this process gives rise to a difference between the number density of DM particles and antiparticles, so that the net abundance survives after the PT.

We assume the wall is planar, perpendicular to the $z$-axis, and moving with the bubble wall velocity $v_w$ in the negative $z$ direction. Also, assuming the system has reached a steady state and is translation invariant in $x$ and $y$, one can write the Liouville operator of phase space distribution function $f_N$ for DMs in the wall frame as 
\begin{equation}
\mathbf{L}[f_N]=\frac{df_N}{dt}=\frac{p_{z}}{E} \frac{\partial f_{N}}{\partial z}+F\frac{\partial f_{N}}{\partial p_{z}}.
\end{equation}
$ F$ is the semiclassical force which is given by \cite{Cline:2020jre,Cline:2000nw,Fromme:2006wx}
\begin{equation}\label{force}
	F=-\frac{m_N(z)}{E}\frac{\partial m_N}{\partial z}\pm s\frac{m_N(z)}{E E_z}\left(\frac{\partial m_N}{\partial z}\frac{\partial \theta}{\partial z}+\frac{m_N(z)}{2}\frac{\partial^2 \theta}{\partial z^2}-\frac{m_N^2(z)}{2E^2}\frac{\partial m_N}{\partial z}\frac{\partial \theta}{\partial z}\right) 
\end{equation}
where
\begin{equation}
	E_z^2=E^2-\mathbf{p}^2_{\|}=m_N^2+p_z^2
\end{equation}
and $\mathbf{p}_{\|} $ denotes the momentum parallel to the wall. The positive sign $(+)$ is assigned to particles and $(-)$ to antiparticles. Also, boosting to the frame in which $\mathbf{p}_{\|}=0 $ for helicity eigenstates, $s=\pm 1$ \cite{Cline:2020jre}.  The last three terms of (\ref{force}) are CP-violating sources during the PT and originated from the complex mass term
\begin{equation}
\hat{m}_N(z)=y_i\phi(z)/\sqrt{2}\exp[i\theta(z)]=m_N(z)\exp[i\theta(z)].
\end{equation} 
Based on the bubble profile found in the previous section, we solve in the following the Boltzmann equation in $T_n$ units and hence we can model the $z$-dependent solution by
\begin{equation}
	\phi(z)=\frac{A}{2}\left(1+\tanh[B z]\right)
\end{equation}
where $A=7.2$, and $ B=3.9$. Moreover, we consider the CP violating phase as \cite{Bruggisser:2017lhc}
\begin{equation}
\theta(z)=\arctan\left(\frac{\Delta\theta}{2}\left[1+\tanh (\frac{z}{l_w})\right]\right).
\end{equation}
Using the ansatz $f_N=\mathcal{A}(z,p_z)\exp(-E^{p}/T) $ for the distribution function \cite{Baker:2019ndr}, we describe the deviation from equilibrium by $ \mathcal{A}$ which also includes the chemical potential. The energy of a particle in the plasma frame $E^{p} $ is related to the one in the wall frame as
\begin{equation}
	E_{p}^{p}=\gamma_w(E-v_wp_z),~~~~\gamma_w=\frac{1}{\sqrt{1-v_w^2}}. 
\end{equation}
Integrating over $ p_x$ and $ p_y$, for RH chirality the Liouville operator would be
\begin{equation}\label{lio}
	\begin{aligned}
		g_{N}& \int \frac{d p_{x} d p_{y}}{(2 \pi)^{2}} \mathbf{L}\left[f_{N}\right] \sim \\
		\sim& \left[\left(\frac{p_{z}}{m_{N}} \frac{\partial}{\partial z}-\left(\frac{\partial m_{N}}{\partial z}\pm\left(\frac{1}{2m_N}\frac{\partial m_N}{\partial z}\frac{\partial \theta}{\partial z}+\frac{1}{2}\frac{\partial^2 \theta}{\partial z^2}\right) \right) \frac{\partial}{\partial p_{z}}-\left(\frac{\partial m_{N}}{\partial z}\right) \frac{v_{w}}{T} \right) \mathcal{A}_{\pm}\left(z, p_{z}\right)\right]\\	
		\times&\frac{g_{N} m_{N} T}{2 \pi} \exp\left(\frac{v_{w} p_{z}-\sqrt{m_{N}^2+p_z^2}}{T}\right)
	\end{aligned}
\end{equation}
where because of the friction effects produced from reflected and penetrated (anti)particles, we used non-relativistic bubble wall velocities, so that $ E^p\simeq E-v_w p_z$. According to the Boltzmann equation, we have $\mathbf{L}\left[f_{N}\right]=\mathbf{C}\left[f_{N}\right] $, where the integration over the collision term is given by \cite{Baker:2019ndr}
\begin{equation}\label{coll}
	g_{N} \int \frac{d p_{x} d p_{y}}{(2 \pi)^{2}} \mathbf{C}\left[f_{N}\right] =-g_{N}^2 \left[\mathcal{A}_{\pm}\left(z, p_{z}\right)-1\right] \int \frac{d p_{x} d p_{y}}{(2 \pi)^{2} 2 E_{p}^{p}} \frac{d^3 q}{(2 \pi)^{3} 2 E_{q}^{p}} 4 \tilde{F} \sigma \exp\left(-\frac{E_{p}^{p}+E_{q}^{p}}{T}\right) 
\end{equation}
where $\sigma$ is the spin-averaged cross section of such processes $ N_R(p^p)+N_R(q^p)\rightarrow \phi(k^p)+\phi(l^p)$ and $ \tilde{F}=\sqrt{((p+q)^2-2m_N^2)^2-4m_N^4}/2$. To solve the Boltzmann equation, we use the method explained in \cite{Baker:2019ndr} and rewrite the equation as
\begin{equation}
a_{\pm}\left(z, p_{z}\right) \frac{\partial \mathcal{A}_{\pm}}{\partial z}+b_{\pm}\left(z, p_{z}\right) \frac{\partial \mathcal{A}_{\pm}}{\partial p_{z}}=c_{\pm}\left(\mathcal{A}_{\pm}, z, p_{z}\right)
\end{equation}
where
\begin{equation}
\frac{d z(\lambda)}{d \lambda}=a_{\pm}\left(z, p_{z}\right), ~~~~~\frac{d p_{z}(\lambda)}{d \lambda}=b_{\pm}\left(z, p_{z}\right),~~~~~
\frac{d \mathcal{A}_{\pm}\left(z(\lambda), p_{z}(\lambda)\right)}{d \lambda}=c_{\pm}\left(\mathcal{A}_{\pm}(\lambda), z(\lambda), p_{z}(\lambda)\right)
\end{equation}
and $\lambda$ parameterizes a given curve on the ($z$, $p_z$) plane. Therefore, from Eqs.\ (\ref{lio}, \ref{coll}), imposing appropriate boundary conditions, we first obtain $ z(\lambda)$ and $p_{z}(\lambda) $, then we find $\mathcal{A}_{\pm} $ for a region of $z$ and $p_z$ values.

\begin{figure}
	\includegraphics[scale=0.5]{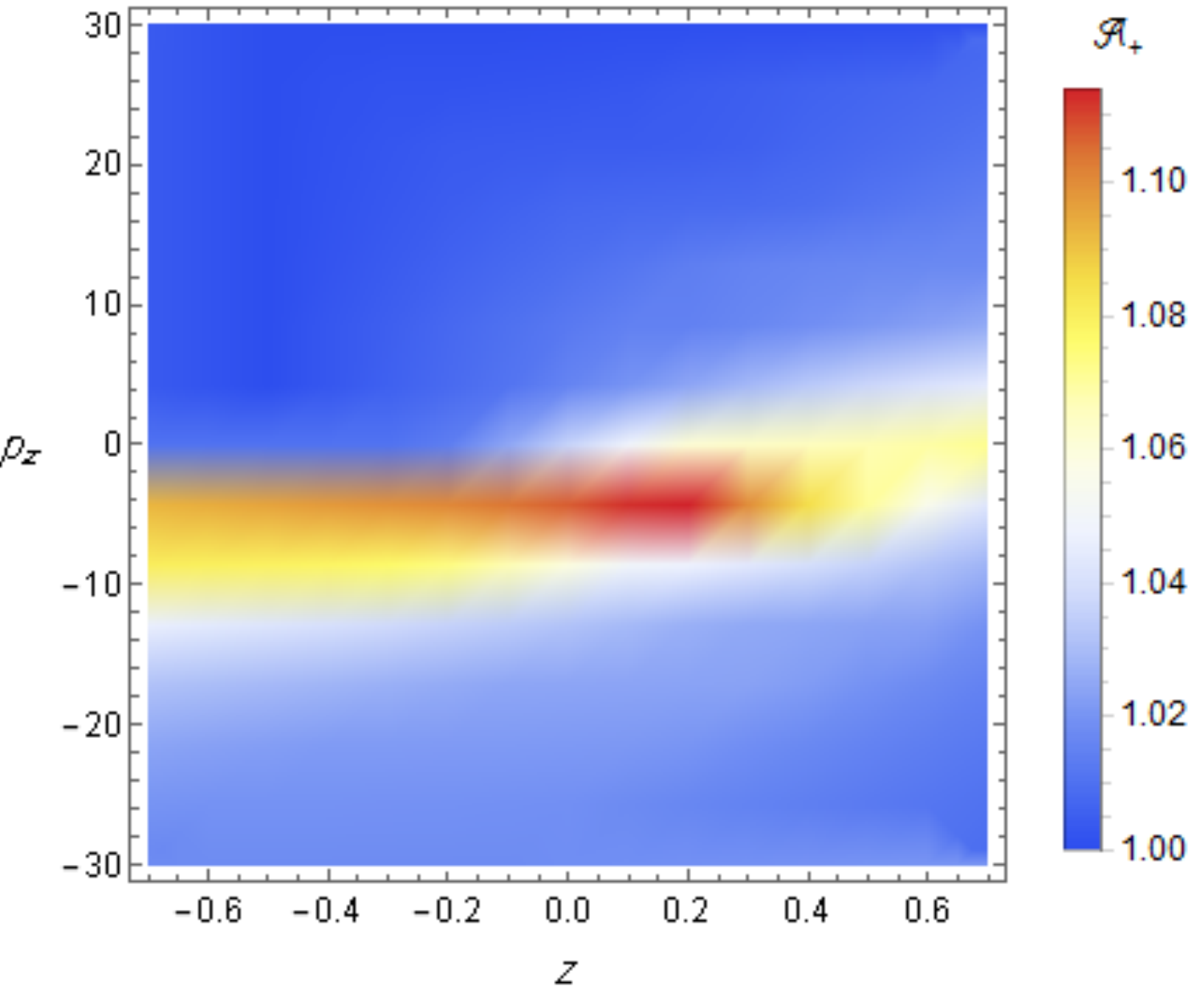}
	\includegraphics[scale=0.5]{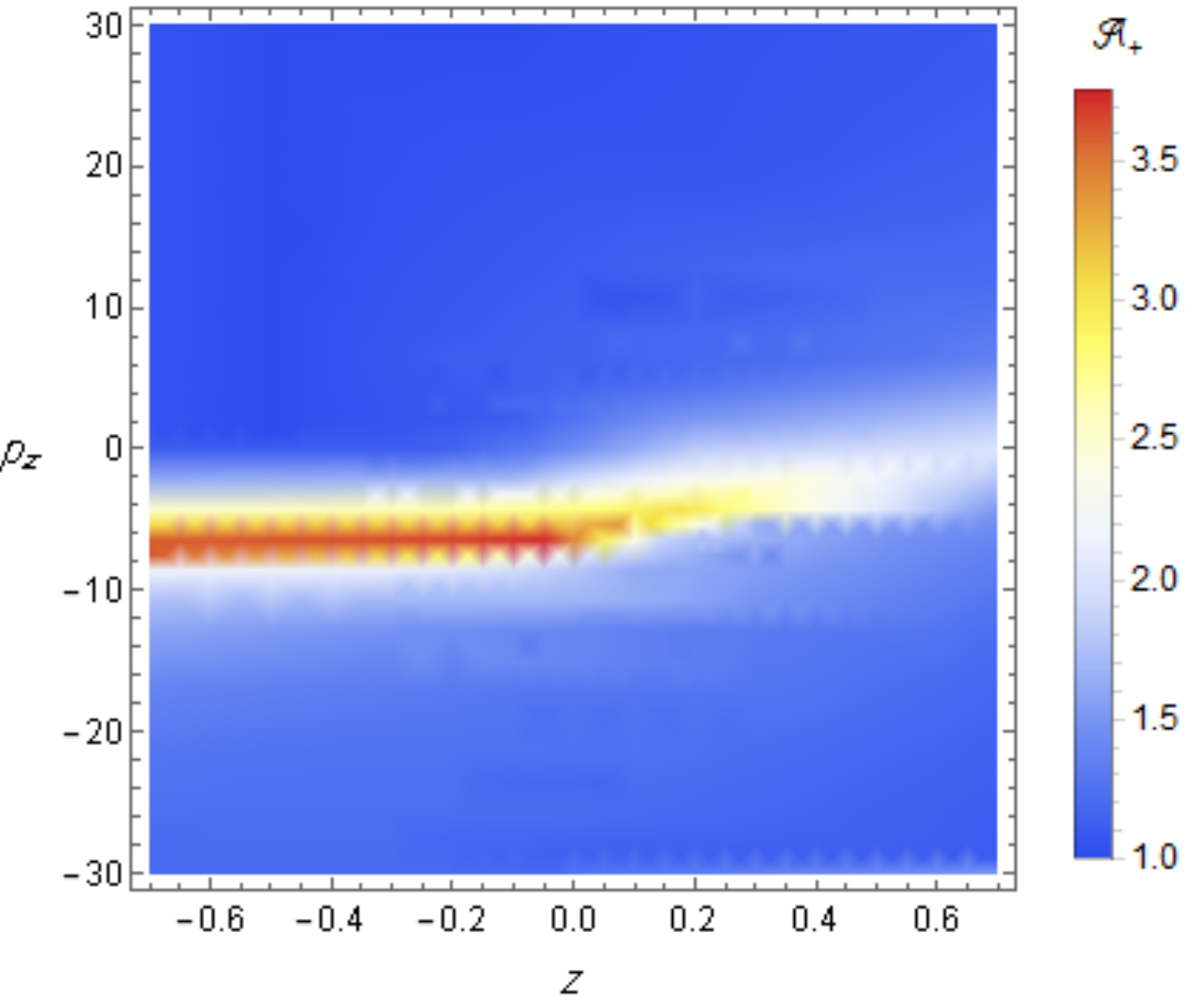}
	\caption{For two bubble velocities, $ v_w=0.01$ (left figure) and $ v_w=0.1$ (right figure), the factor $\mathcal{A}_{+} $ for DM particles is displayed. Since to obtain the observed relic abundance, it requires $ \Delta\theta\ll 1$ and hence since $\mathcal{A}_{+} $ and $\mathcal{A}_{-} $ are displayed almost the same,  $\mathcal{A}_{+} $ is only shown.}
	\label{fd}
\end{figure}

For (anti)particles outside the bubble traveling with $p_z>0$, the boundary condition is set as
\begin{equation}
\lim _{z \rightarrow-\infty} \mathcal{A}_{\pm} \rightarrow 1,
\end{equation}
ensuring the equilibrium phase space distribution function far away from the bubble wall. Using the obtained curves of ($z\rightarrow-\infty$, $p_z>m_{N}^{\mathrm{in}}$), for (anti)particles deep inside the bubble ($z\rightarrow\infty$) with $ p_z<0$ and similar dynamics, we set the boundary condition as \cite{Baker:2019ndr}
\begin{equation}
\lim _{z \rightarrow \infty} \mathcal{A}_{\pm}(p_z)=\lim _{z \rightarrow \infty} \mathcal{A}_{\pm}(-p_z).
\end{equation}
Interpolating between different solutions obtained for several curves, we find $\mathcal{A}_{\pm}$ as displayed in Fig.\ (\ref{fd}). Finally, integrating over $p_z$ at deep inside the bubble, we obtain the net number density $ \Delta n_N/T_n^3$ in $T_n$ units, where $\Delta n_N\equiv n_N^+-n_N^-$. We take $ v_w=0.01$ in the following calculations, however, for other values, for example $ v_w=0.1$, the final result can be equivalently obtained by slightly different values of parameters.

The DM relic abundance can be found in the asymmetric case by the following relation\footnote{During the PT, L-violating processes with CP-violating effects, present during this period, occur inside the bubbles, thus the net number density is used in the DM relic abundance relation.} 
\begin{equation}
h^{2}\Omega_{\mathrm{DM}} \simeq \frac{m^{\mathrm{in}}_N \Delta n_N ~g_{* 0} T_0^3}{3 M_{\mathrm{pl}}^{2}\left(H_{0} / h\right)^{2}g_{*} T_n^3}
\end{equation}
where $g_{* 0}=3.9$, $T_0\simeq 0.235\,\mathrm{meV}$, $g_*\sim 110$ and the present Hubble parameter $ H_0=100\,h~\mathrm{km}\,\mathrm{sec}^{-1}\,\mathrm{Mpc}^{-1} $. As a result, the observed relic abundance, $h^{2}\Omega_{\mathrm{DM}} \simeq 0.12 $ \cite{Planck:2018vyg}, can be obtained for, e.g., $y_1=\sqrt{2}$ or $ m^{\mathrm{in}}_N\simeq 66.3\,\mathrm{PeV}$ and from the attained $ \Delta n_N/T_n^3$ which is determined by $ \Delta\theta\sim 10^{-12}$.

Furthermore, analogous to the analytic approach of \cite{Baker:2019ndr}, estimated the DM number density proportional to its equilibrium abundance inside the bubble $n_{\mathrm{eq}}/4$ where
\begin{equation}
n_{\mathrm{eq}}= \frac{g_{N}\left(m_{N}^{\mathrm{in}} T_{n}\right)^{3 / 2}}{(2 \pi)^{3 / 2}} e^{-m_{N}^{\mathrm{in}} / T_{n}},
\end{equation}
here, considering the net number density as $ \Delta n_N\sim \delta\theta\, n_{\mathrm{eq}}$, with $ T_n\simeq 9.19\,\mathrm{PeV}$ and $m^{\mathrm{in}}_N\simeq 66.3\,\mathrm{PeV}$, one can obtain the observed relic abundance for $\delta\theta\simeq 0.19\times 10^{-12}$.

\section{Phenomenology}
\label{phen}
In this section we study phenomenological consequences of the model.
\subsection{Axion couplings}
The Lagrangian of the QCD axion as a dynamical solution for the strong CP problem is given such that the axion effective interaction with gluons and its vev give rise to cancellation of the $ \bar{\theta}$ term. Along this direction, the anomalous PQ current of the axial $ U(1)_{\mathrm{PQ}}$ symmetry is given as follows
\begin{equation}
	\partial^{\mu} J_{\mu}^{\mathrm{PQ}}=\frac{g_{s}^{2} \mathcal{N}}{16 \pi^{2}} G \widetilde{G}+\frac{e^{2} \mathcal{E}}{16 \pi^{2}} F \widetilde{F}
\end{equation} 
where $ g_s$ is the $ \mathrm{SU(3)}_{c}$ strong coupling, $ e$ is the electric charge, $ F_{\mu\nu}$ is the photon field strength, $\mathcal{N}$ and $ \mathcal{E}$ are  the QCD and electromagnetic anomaly coefficients, respectively. Due to the spontaneous PQ symmetry breaking, the effective interactions of the axion are expressed as
\begin{equation}
	\mathcal{L}_{a} \supset \frac{g_{s}^{2}}{32 \pi^{2} f_a}a G \widetilde{G}+\frac{g_{a\gamma}}{4} a F \widetilde{F}-\frac{c^0_f}{2f_{a}} \partial_{\mu} a \bar{f}\gamma^{\mu}\gamma_5 f-\frac{c^0_{N_R}}{2f_{a}} \partial_{\mu} a \overline{N}_R \gamma^{\mu} N_R
\end{equation}
where
\begin{equation}
	g_{a\gamma}=\frac{e^2}{8\pi^2f_a}\frac{\mathcal{E}}{\mathcal{N}}-\frac{e^2}{8\pi^2f_a}\left(\frac{2}{3}\frac{4m_d+m_u}{m_d+m_u}\right)
\end{equation}
\begin{equation}
	c^0_f=\frac{X_{f_R}-X_{f_L}}{2\mathcal{N}}=\frac{X_{H_f}}{2\mathcal{N}},~~~~~~~c^0_{N_R}=\frac{X_{\Phi}}{2\mathcal{N}}
\end{equation}
\begin{equation}
	\mathcal{N}=\frac{n_g}{2}\left(X_{H_u}+X_{H_d}\right)
\end{equation}
\begin{equation}
	\mathcal{E}=n_g\left(3\left(\frac{2}{3}\right)^2 X_{H_u}+3\left(-\frac{1}{3}\right)^2 X_{H_d}+\left(-1\right)^2 X_{H_d}\right)
\end{equation}
$ n_g$ is the number of SM fermion $f$ generation, $f_a=v_a/(2\mathcal{N})$, $m_u$ and $m_d$ are $u$ and $d$ quark masses, respectively. The cosmological Domain Wall (DW) problem \cite{Zeldovich:1974uw} can be avoided if $ N_{\mathrm{DW}}\equiv2\mathcal{N}=1$ \cite{DiLuzio:2020wdo}. As explained in Section \ref{model}, $X_{H_u}+X_{H_d}=X_{\Phi} $. Therefore, according to Table \ref{t1}, $\mathcal{N}=1/2 $ and thereby $f_a=v_a$ and $N_{\mathrm{DW}}=1 $ can be fulfilled if one family is charged under the PQ symmetry.

Moreover, depending on the fermion type, $ \tan\theta\in[0.25, 170] $ \cite{DiLuzio:2020wdo}, thus we obtain $ \mathcal{E}=4/3$, $ c_u^0=\cos^2\theta$, $ c_d^0=\sin^2\theta$, $ c_{N_R}^0=1$ from which other couplings including the axion-pion coupling can de determined.

\subsection{Gravitational wave signals}
Regarding growing efforts and progresses on the GW direct detection, cosmological PTs, specially first order PTs as a source for the GW radiation, are important events that should be studied \cite{Mazumdar:2018dfl,Ahmadvand:2017xrw,Ahmadvand:2017tue,Abedi:2019msi,Ahmadvand:2020fqv}.

During cosmological first order PTs and bubble evolution processes, three sources for the GW generation have been proposed: bubble collision, sound waves and magnetohydrodynamic  (MHD) turbulence \cite{Kosowsky:1992rz,Kamionkowski:1993fg,Kosowsky:2001xp,Hindmarsh:2015qta}. The GW energy density spectrum can be characterized by some of important PT quantities including the bubble wall velocity, the released latent heat, and duration of the PT, calculated at the nucleation temperature.\footnote{Since we do not differentiate between the temperature at which GWs are generated and the nucleation temperature in this prompt PT, we calculate the parameters at $T_n$.}

The parameter which is associated with the latent heat and appears in the GW energy density computation is the ratio of the vacuum energy density to the thermal energy density,
\begin{equation}\label{al}
	\alpha =\frac{\epsilon}{\frac{\pi^2}{30}g_*T_n^4},~~~~~~\epsilon=\left(\Delta V_{t}(T)-T\frac{d\Delta V_{t}(T)}{dT}\right)\Bigg|_{T=T_n},
\end{equation}
where $ \Delta V_{t}(T_n)=V_{t}[0, T_n]-V_{t}[v_t (T_n), T_n] $ and $ v_t (T_n)$ is the true vacuum at $ T_n$.
There is a critical value of  $ \alpha $, denoted by $\alpha_{\infty} $, such that for $ \alpha>\alpha_{\infty}$ bubbles can run away \cite{Caprini:2015zlo}, where
\begin{equation}\label{inf}
	\alpha_{\infty}=\frac{30}{24 \pi ^2}\frac{\sum _i n_i \Delta m_i ^2}{g_* T_n^2},
\end{equation}
$ n_i\,(n_i/2) $ is the number of degrees of freedom for boson (fermion) species, and $ \Delta m_i^2 $ is the squared mass difference of particles between the broken and symmetric phase at the nucleation temperature. Another key parameter related to the inverse of PT duration is calculated by
\begin{equation}\label{beta}
	\frac{\beta}{H_*}=T_n\frac{d}{dT}\left(\frac{S_3(T)}{T}\right)\Bigg |_{T_n}. 
\end{equation}
Based on the analysis described in Section \ref{first}, we found $ T_n\simeq 9.19\,\mathrm{PeV}$. At this temperature we obtain $ \alpha\simeq 0.55$, $\alpha_{\infty}\simeq 0.6 $. In addition, from Eq.\ (\ref{beta}) and the obtained bounce action we find $\beta/H_*\simeq 246$.

In the non-runaway case where $ \alpha<\alpha_{\infty}$, dominant contributions to GWs come from sound waves and MHD turbulence, i.e., $ h^2\Omega (f)\simeq h^2\Omega _{\mathrm{sw}}+h^2\Omega _{\mathrm{tu}} $ where \cite{Hindmarsh:2015qta,Caprini:2009yp}
\begin{equation}\label{sps}
h^2\Omega _{\mathrm{sw}}(f)=2.65\times 10^{-6}\Big(\frac{H_*}{\beta}\Big)\Big(\frac{\kappa _{\mathrm{sw}} \alpha}{1+\alpha}\Big)^2\Big(\frac{100}{g_*}\Big)^{\frac{1}{3}} v_{w}~ S_{\mathrm{sw}}(f),
\end{equation}
\begin{equation}\label{spt}
	h^2\Omega _{\mathrm{tu}}(f)=3.35\times 10^{-4}\Big(\frac{H_*}{\beta}\Big)\Big(\frac{\kappa _{\mathrm{tu}} \alpha}{1+\alpha}\Big)^{\frac{3}{2}}\Big(\frac{100}{g_*}\Big)^{\frac{1}{3}} v_{w}~ S_{\mathrm{tu}}(f),
\end{equation}
where the spectral shapes are given by \cite{Caprini:2015zlo},
\begin{eqnarray}
	S_{\mathrm{sw}}(f)&=&\Big(\frac{f}{f_{\mathrm{sw}}}\Big)^3\Big(\frac{7}{4+3(\frac{f}{f_{\mathrm{sw}}})^{2}}\Big)^{\frac{7}{2}}, \\
	S_{\mathrm{tu}}(f)&=&\frac{(\frac{f}{f_{\mathrm{tu}}})^3}{(1+\frac{f}{f_{\mathrm{tu}}})^{\frac{11}{3}} (1+\frac{8\pi f}{h_*})},
\end{eqnarray}
with
\begin{equation}
	h_*=16.5\times 10^{-2} [\mathrm{Hz}]\Big(\frac{T_n}{\mathrm{PeV}}\Big)\Big(\frac{g_*}{100}\Big)^{\frac{1}{6}}.
\end{equation}
The red-shifted peak frequencies in the spectral shapes are given by
\begin{equation}\label{fst}
	f_{\mathrm{sw}}=19\times 10^{-2}[\mathrm{Hz}]  \Big(\frac{1}{v_w}\Big)\Big(\frac{\beta}{H_*}\Big)\Big(\frac{T_n}{\mathrm{PeV}}\Big)\Big(\frac{g_*}{100}\Big)^{\frac{1}{6}},
\end{equation}
\begin{equation}
	f_{\mathrm{tu}}=27\times 10^{-2}[\mathrm{Hz}] \Big(\frac{1}{v_w}\Big)\Big(\frac{\beta}{H_*}\Big)\Big(\frac{T_n}{\mathrm{PeV}}\Big)\Big(\frac{g_*}{100}\Big)^{\frac{1}{6}}.
\end{equation}
In this case, the bubble wall velocity reaches to a subluminal value and depending on the velocity, the efficiency factor for the conversion of the latent heat to the plasma motion would be different. We study the effect of various bubble wall velocities and combustion modes on the GW signals. For small velocities, $ v_w\ll c_s=1/\sqrt{3}$,  $ v_w= c_s$ which is the case of transition from subsonic to supersonic deflagrations, and the large velocity limit, $ v_w\rightarrow 1$, this efficiency factor is expressed respectively as \cite{Espinosa:2010hh}
\begin{equation}
\kappa ^A_{v}=\frac{6.9\alpha~v_w^{6/5}}{1.36-0.037\sqrt{\alpha}+\alpha},~~~~~ v_w\ll c_s,
\end{equation}
\begin{equation}
\kappa ^B_{v}=\frac{\alpha^{2/5}}{0.017+(0.997+\alpha)^{2/5}},~~~~~v_w= c_s,
\end{equation}
\begin{equation}
\kappa ^C_{v}=\frac{\alpha}{0.73+0.083\sqrt{\alpha}+\alpha},~~~~~v_w\rightarrow 1.
\end{equation}
Moreover, for subsonic deflagrations $ v_w\lesssim c_s$, Jouguet detonations, and detonations with $ v_w\gtrsim v_J$, the efficiency factor is given by the following fits \cite{Espinosa:2010hh}
\begin{equation}
\kappa_{v}\left(v_{w}\lesssim c_s\right)=\frac{c_s^{11 / 5} \kappa_{v}^{A} \kappa_{v}^{B}}{\left(c_s^{11 / 5}-v_{w}^{11 / 5}\right) \kappa_{v}^{B}+v_{w}\,c_{s}^{6 / 5} \kappa_{v}^{A}},
\end{equation}
\begin{equation}
\kappa^J_{v}=\frac{\sqrt{\alpha}}{0.135+\sqrt{0.98+\alpha}}, \quad v_w=v_{J}=\frac{\sqrt{2\alpha/3+\alpha^{2}}+\sqrt{1/3}}{1+\alpha},
\end{equation}
\begin{equation}\label{effi}
\kappa_{v}\left(v_{w} \gtrsim  v_{J}\right)=\frac{\left(v_{J}-1\right)^{3}\left(v_{J} / v_{w}\right)^{5 / 2} \kappa_{v}^{J} \kappa_{v}^{C}}{\left[\left(v_{J}-1\right)^{3}-\left(v_{w}-1\right)^{3}\right] v_{J}^{5 / 2} \kappa_{v}^{J}+\left(v_{w}-1\right)^{3} \kappa_{v}^{C}}.
\end{equation}
The fraction of plasma motion in turbulence, $\varepsilon=\kappa _{\mathrm{tu}}/\kappa _v $, can be set as $ \varepsilon =0.05 $ \cite{Hindmarsh:2015qta}, so that the main source contributing to the GW energy density from the plasma motion is the source of sound waves, $ \kappa _{\mathrm{sw}}=(1-\varepsilon)\kappa _v $. 

\begin{figure}[tbp]
	\centering
	\includegraphics[scale=0.5]{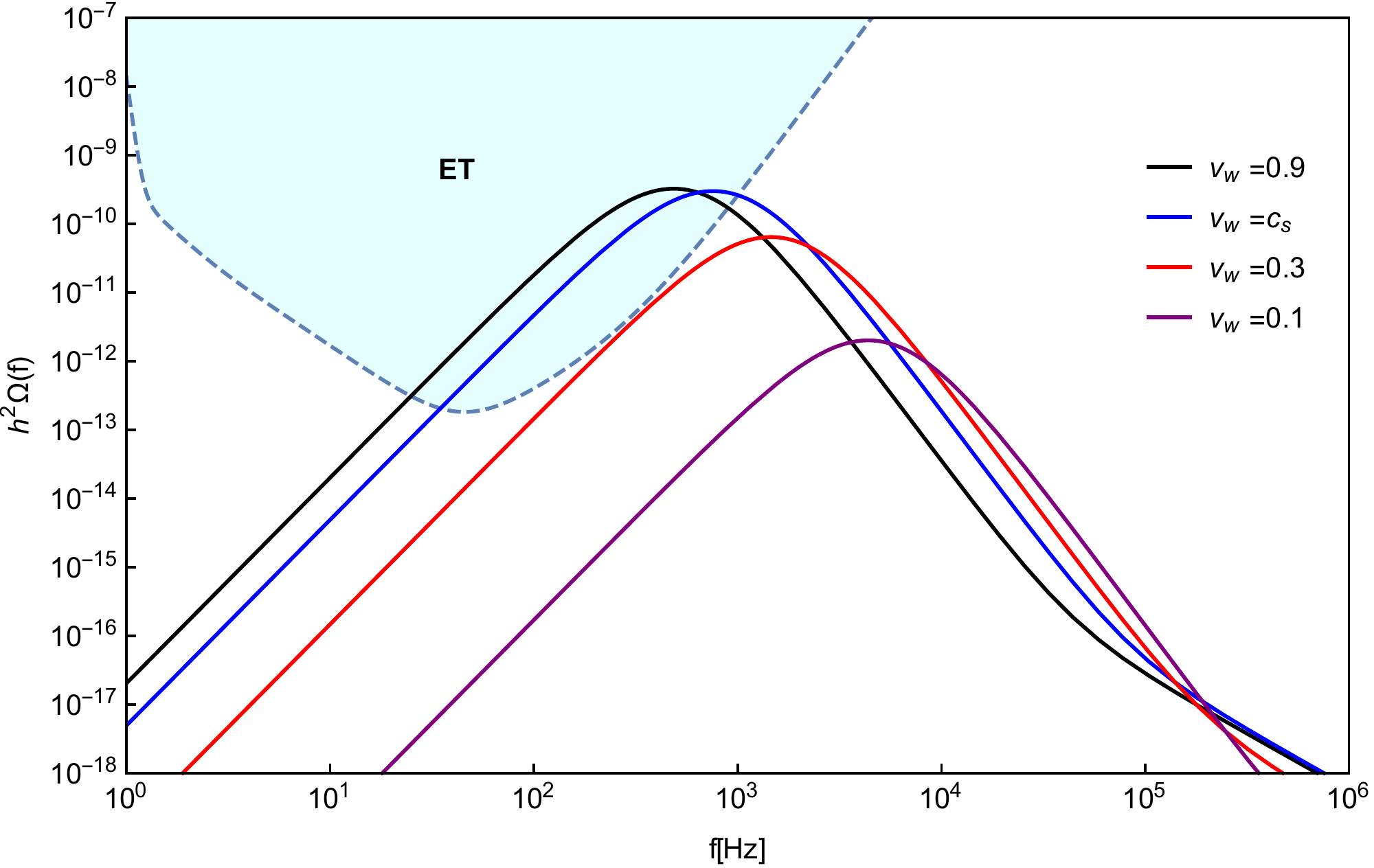}
	\caption{GW energy density spectra of the first order PQ PT with $ f_a=100\,\mathrm{PeV}$ for different bubble wall velocities and combustion regimes are displayed. For $ v_w=0.9$, we use Eq.\ (\ref{effi}) for the efficiency factor. The dashed curve is the approximate prospective sensitivity curve of ET \cite{Hild:2008ng,Sathyaprakash:2012jk}. We show that the GW signals may be within the reach of this planned GW detector.}
	\label{f3}
\end{figure}

Eventually, substituting key parameters in Eqs.\ (\ref{sps}, \ref{spt}), we find the GW spectrum for different cases of the bubble wall velocity. As shown in Fig.\ (\ref{f3}), the GWs of the first order PQ PT can be within the reach of ongoing third generation ground-based detectors such as Einstein Telescope (ET) \cite{Hild:2008ng,Sathyaprakash:2012jk}

\subsubsection{Gravitational wave signals within a heavy axion case}
It is also interesting to study the possibility of the so-called visible axion models in which the PQ symmetry breaking scale can be $ f_a\sim 1-100\,\mathrm{TeV}$ \cite{Fukuda:2015ana}. 

We here only explore the detectability of possible GWs from the first order PQ PT in a heavy axion model. Such class of models can be justified by considering an additional $ \mathbb{Z}_{\mathbf{N}} $ mirror symmetry with $\mathbf{N} $ mirror worlds, so that with additional confining sectors the axion potential would change, $ (a/f_a +\bar{\theta}_{\mathrm{eff}})G_k\widetilde{G}_k $ \cite{Rubakov:1997vp,Berezhiani:2000gh,Hook:2014cda,Fukuda:2015ana}. Consequently, in this case the axion mass would be \cite{Fukuda:2015ana}
\begin{equation}
m_{a} \simeq \frac{\sqrt{z^{\prime}}}{1+z^{\prime}} \frac{f_{\pi^{\prime}} m_{\pi^{\prime}}}{f_{a}}
\end{equation}
where  $ m_{\pi^{\prime}}$ and $f_{\pi^{\prime}}$ are the pion mass and its decay constant in the mirror sectors and $z^{\prime}=m_{u^{\prime}}/m_{d^{\prime}}$. For example taking $z^{\prime}=z\sim 0.5$, $m_{\pi^{\prime}}=135\,\mathrm{GeV}$, $f_{\pi^{\prime}}=93\,\mathrm{GeV}$ and $ f_a=100\,\mathrm{TeV}$, the axion mass will be $m_a\sim 60\,\mathrm{MeV}$.

\begin{figure}[tbp]
	\centering
	\includegraphics[scale=0.5]{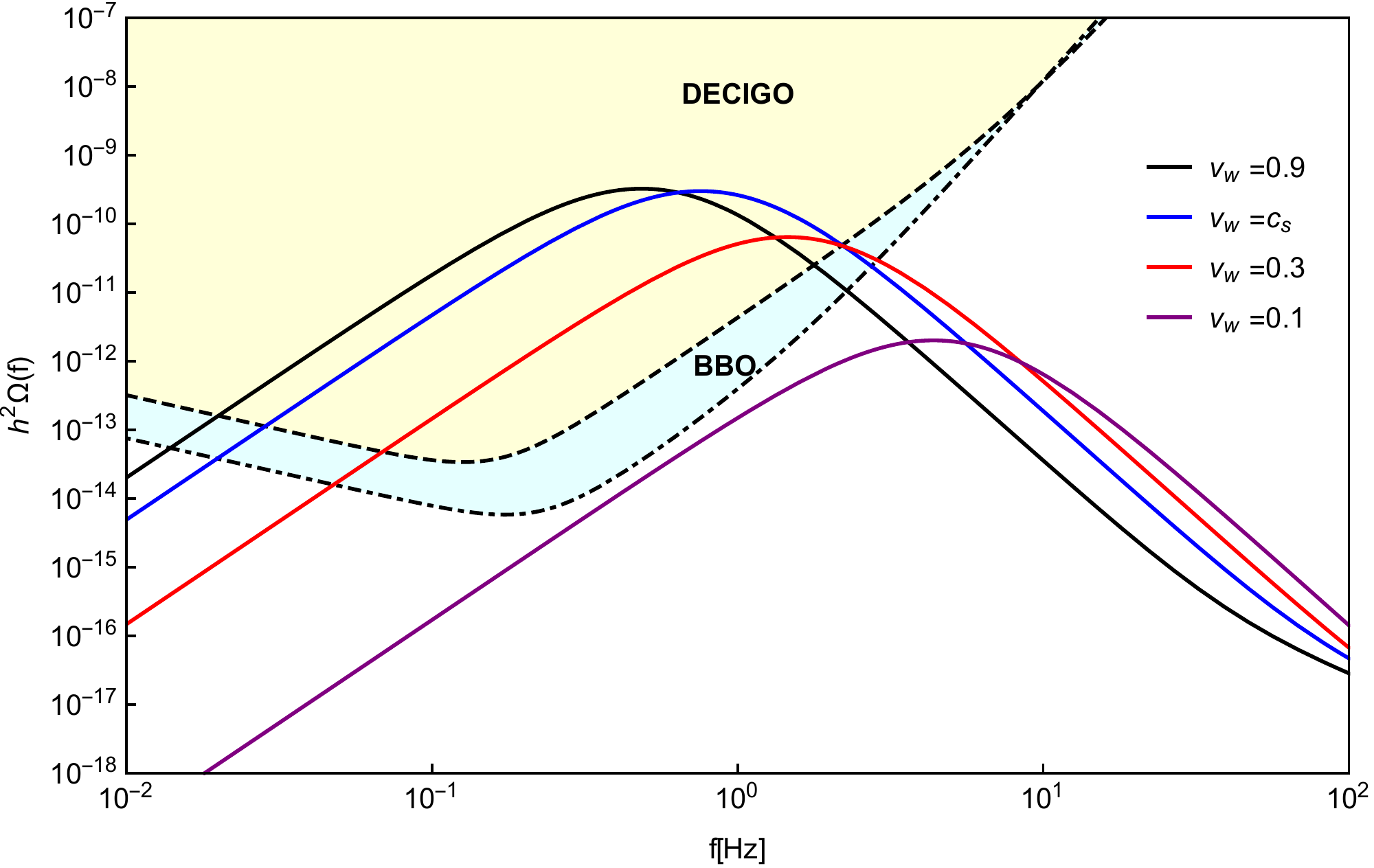}
	\caption{GW energy density spectra of the first order PQ PT with $ f_a=100\,\mathrm{TeV}$ for different bubble wall velocities are displayed. Dashed curves are the expected sensitivity curves of proposed space-based DECIGO and BBO detectors \cite{Yagi:2011wg,Kudoh:2005as}, which may probe the generated GW signals.}
	\label{f4}
\end{figure}

Connecting the mirror worlds by the axion \cite{DiLuzio:2021pxd}, in the UV axion model the PQ scalar can be transformed as $ \phi\rightarrow \exp (2\pi i/\mathbf{N}) \phi$. Thus, considering $ f_a=100\,\mathrm{TeV}$, the PT parameters can be retained at $ T_n\simeq 9.19\,\mathrm{TeV}$. As a result, as displayed in Fig.\ (\ref{f4}), in this case the GW signals of the PQ PT can be probed by the future space-based interferometers including  DECIGO and BBO \cite{Yagi:2011wg,Kudoh:2005as}.

We leave further investigations on these models for another future work.

\subsection{Effective dark matter interactions}
\begin{figure}
	\centering
		\centering
		\includegraphics[scale=1.05]{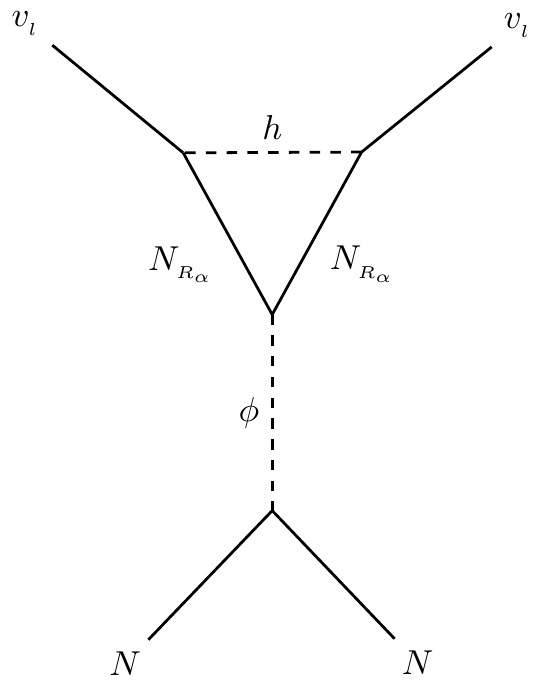} 
		\hspace{0.5cm}
		\includegraphics[scale=1.1]{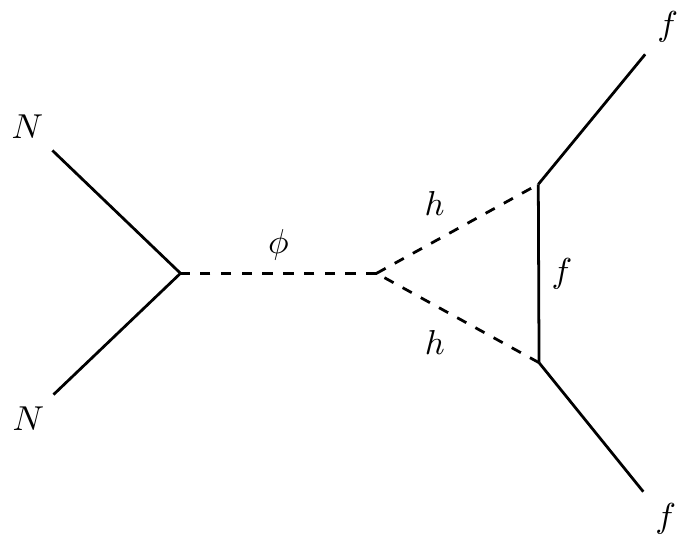}
	\caption{Left: The DM annihilation to SM neutrinos is shown, where $h$ is the Higgs field. Right: The DM-SM fermion scattering is shown. }
	\label{fdm}
\end{figure}

Depending on DM interactions with the visible sector, a model can be also tested by direct detection experiments and indirect measurements. In this section we illustrate some of these processes, particularly focusing on the DM-SM neutrino interaction. However, we first consider elastic $ N\,e\rightarrow N^{\prime}\,e^{\prime}$ scattering (Fig\. (\ref{fdm})) and explore the electron recoil energy.
 
DMs move at non-relativistic velocities, $v_D\sim 10^{-3}$, in typical DM halos \cite{Smith:2006ym}. Considering these speeds for the PeV-scale DM, one can calculate the transfer recoil energy as follows \cite{Kannike:2020agf}
\begin{equation}
E_r\equiv E_e-E_{e^{\prime}}\simeq 2m_e v_D (v_D-v_e),
\end{equation}
hence in the lab frame for $v_e=0$ and $m_e\simeq 0.5\,\mathrm{MeV}$, the electron recoil energy would be $E_r\simeq \mathrm{eV}$. Therefore, such non-relativistic massive DMs cannot explain the excess events in a $1-7\,\mathrm{keV}$ recoil energy reported by the XENON1T experiment \cite{XENON:2020rca}.\footnote{For an explanation for these events, see for example \cite{Shakeri:2020wvk}.}

However, the heavy DMs can be considered as a source for producing ultra-high energy neutrinos \cite{Arguelles:2019ouk}. To show this process according to the model (Fig.\ (\ref{fdm})), we consider the annihilation of DM to SM neutrinos $N(p)\,N(q)\rightarrow v_l(k)\,v_l(g)$. In the center of mass frame, for the non-relativistic DM, $v_D\sim 10^{-3}$, we have $|\mathbf{p}|=|\mathbf{q}|\sim m_N v_D$ and hence the energy of incoming particles $E\sim m_N$. Also, for $m_{v_l}\sim 0$ and due to the energy and momentum conservation, $E_{v_l}\sim|\mathbf{k}|\sim m_N$. Finally, calculating the amplitude of the process, $\mathcal{M}$, we can find the cross section as follows
\begin{equation}\label{cross}
\sigma\sim\frac{|\mathbf{k}|\, |\mathcal{M}|^2}{64\pi E^2\, |\mathbf{p}|}\sim 10^{-48}\,\mathrm{cm}^2\left(\frac{\mathcal{Y}}{10^{-2}}\right)^2\left(\frac{m_N}{1\,\mathrm{PeV}}\right)^{-2} 
\end{equation}
where $\mathcal{Y}\equiv y_1y_{\alpha} y_{N_{\alpha}}^2$ and we take $ m_N=y_1f_a/\sqrt{2}\gg m_{\phi}$. Therefore, as obtained from Eq.\ (\ref{cross}), for $E_{v_l}\sim m_N=1\,\mathrm{PeV}$, the cross section is compatible with the bound on DM annihilation processes \cite{Chianese:2021htv}.

\section{Conclusion}
\label{con}
The bubble filtration mechanism provides a setup which allows DM masses above $100\,\mathrm{TeV}$, the GK bound constrained the DM mass within the thermal freeze-out mechanism. In this work, using the filtering mechanism, we have presented an asymmetric DM scenario during the PQ PT through which DMs can naturally acquire these large masses. Based on a QCD axion model extended by chiral neutrinos, where one of the flavors plays the role of DM, we find the one-loop finite temperature effective potential and show that the PT can be first order within the parameter space of the theory. We obtain the profile of bubbles nucleated during the PT at the PQ symmetry breaking scale around $100\,\mathrm{PeV}$. Relying on the lepton number violating interaction of RH neutrinos, we have shown an asymmetry between heavy neutrinos and antineutrinos, provided that a CP-violating source is imposed during the PT. Indeed, we solve the Boltzmann equation numerically, and find the net number density as well as the observed relic abundance of the DM. It is interesting to note that because of CP violation effects varying during the PT, the scenario is not restricted to $m_N\gg T_n$ for obtaining the relic abundance and also the resulting net abundance remains after the PT.

As for resolving the strong CP problem in the model, we have then obtained the QCD axion couplings. Furthermore, calculating the vacuum energy and duration of the PT at the nucleation temperature, we find the energy density spectrum of GWs generated from the PQ PT for different combustion modes. We show that the signals can be detected by the future ground-based detectors such as ET. In particular, we have investigated possible GWs of the first order PQ PT for a class of heavy axion models at $f_a=100\,\mathrm{TeV}$. We show the GW signals in this case can be explored by DECIGO and BBO detectors. Eventually, considering the annihilation of DM to SM neutrinos, we compute the cross section which is consistent with the constraint on DM annihilation processes and show that the interaction can be regarded as a source for the ultra-high energy, PeV scale, neutrinos.

\acknowledgments
I would like to thank Soroush Shakeri for helpful comments and discussions.



\begin{thebibliography}{99}
\bibitem{Bertone:2016nfn}
G.~Bertone and D.~Hooper,
``History of dark matter,''
Rev. Mod. Phys. \textbf{90}, no.4, 045002 (2018)
%doi:10.1103/RevModPhys.90.045002
[arXiv:1605.04909 [astro-ph.CO]].	

\bibitem{Bertone:2004pz}
G.~Bertone, D.~Hooper and J.~Silk,
``Particle dark matter: Evidence, candidates and constraints,''
Phys. Rept. \textbf{405}, 279-390 (2005)
%doi:10.1016/j.physrep.2004.08.031
[arXiv:hep-ph/0404175 [hep-ph]].

\bibitem{Allahverdi:2010rh}
R.~Allahverdi, B.~Dutta and K.~Sinha,
``Cladogenesis: Baryon-Dark Matter Coincidence from Branchings in Moduli Decay,''
Phys. Rev. D \textbf{83}, 083502 (2011)
%doi:10.1103/PhysRevD.83.083502
[arXiv:1011.1286 [hep-ph]].

\bibitem{Ahmadvand:2020izy}
M.~Ahmadvand,
``Matter and dark matter asymmetry from a composite Higgs model,''
Eur. Phys. J. C \textbf{81}, no.4, 358 (2021)
%doi:10.1140/epjc/s10052-021-09159-9
[arXiv:2010.10121 [hep-ph]].

\bibitem{Kolb:1998ki}
E.~W.~Kolb, D.~J.~H.~Chung and A.~Riotto,
``WIMPzillas!,''
AIP Conf. Proc. \textbf{484}, no.1, 91-105 (1999)
%doi:10.1063/1.59655
[arXiv:hep-ph/9810361 [hep-ph]].

\bibitem{Kaplan:2009ag}
D.~E.~Kaplan, M.~A.~Luty and K.~M.~Zurek,
``Asymmetric Dark Matter,''
Phys. Rev. D \textbf{79}, 115016 (2009)
%doi:10.1103/PhysRevD.79.115016
[arXiv:0901.4117 [hep-ph]].

\bibitem{Baker:2019ndr}
M.~J.~Baker, J.~Kopp and A.~J.~Long,
``Filtered Dark Matter at a First Order Phase Transition,''
Phys. Rev. Lett. \textbf{125}, no.15, 151102 (2020)
%doi:10.1103/PhysRevLett.125.151102
[arXiv:1912.02830 [hep-ph]].

\bibitem{Chway:2019kft}
D.~Chway, T.~H.~Jung and C.~S.~Shin,
``Dark matter filtering-out effect during a first-order phase transition,''
Phys. Rev. D \textbf{101}, no.9, 095019 (2020)
%doi:10.1103/PhysRevD.101.095019
[arXiv:1912.04238 [hep-ph]].

\bibitem{Peccei:1977hh}
R.~D.~Peccei and H.~R.~Quinn,
``CP Conservation in the Presence of Instantons,''
Phys. Rev. Lett. \textbf{38}, 1440-1443 (1977).

\bibitem{Peccei:2006as}
R.~D.~Peccei,
``The Strong CP problem and axions,''
Lect. Notes Phys. \textbf{741}, 3-17 (2008)
%doi:10.1007/978-3-540-73518-2\_1
[arXiv:hep-ph/0607268 [hep-ph]].

\bibitem{Raffelt:2006cw}
G.~G.~Raffelt,
``Astrophysical axion bounds,''
Lect. Notes Phys. \textbf{741}, 51-71 (2008)
%doi:10.1007/978-3-540-73518-2\_3
[arXiv:hep-ph/0611350 [hep-ph]].

\bibitem{Arvanitaki:2009fg}
A.~Arvanitaki, S.~Dimopoulos, S.~Dubovsky, N.~Kaloper and J.~March-Russell,
``String Axiverse,''
Phys. Rev. D \textbf{81}, 123530 (2010)
%doi:10.1103/PhysRevD.81.123530
[arXiv:0905.4720 [hep-th]].

\bibitem{Dine:1981rt}
M.~Dine, W.~Fischler and M.~Srednicki,
``A Simple Solution to the Strong CP Problem with a Harmless Axion,''
Phys. Lett. B \textbf{104}, 199-202 (1981).

\bibitem{Zhitnitsky:1980tq}
A.~R.~Zhitnitsky,
``On Possible Suppression of the Axion Hadron Interactions. (In Russian),''
Sov. J. Nucl. Phys. \textbf{31}, 260 (1980).

\bibitem{Baldes:2016gaf}
I.~Baldes, T.~Konstandin and G.~Servant,
``Flavor Cosmology: Dynamical Yukawas in the Froggatt-Nielsen Mechanism,''
JHEP \textbf{12}, 073 (2016)
%doi:10.1007/JHEP12(2016)073
[arXiv:1608.03254 [hep-ph]].

\bibitem{Cline:2020jre}
J.~M.~Cline and K.~Kainulainen,
``Electroweak baryogenesis at high bubble wall velocities,''
Phys. Rev. D \textbf{101}, no.6, 063525 (2020)
%doi:10.1103/PhysRevD.101.063525
[arXiv:2001.00568 [hep-ph]].

\bibitem{Long:2017rdo}
A.~J.~Long, A.~Tesi and L.~T.~Wang,
``Baryogenesis at a Lepton-Number-Breaking Phase Transition,''
JHEP \textbf{10}, 095 (2017)
%doi:10.1007/JHEP10(2017)095
[arXiv:1703.04902 [hep-ph]].

\bibitem{Davidson:2008bu}
S.~Davidson, E.~Nardi and Y.~Nir,
``Leptogenesis,''
Phys. Rept. \textbf{466}, 105-177 (2008)
%doi:10.1016/j.physrep.2008.06.002
[arXiv:0802.2962 [hep-ph]].

\bibitem{Pascoli:2016gkf}
S.~Pascoli, J.~Turner and Y.~L.~Zhou,
``Baryogenesis via leptonic CP-violating phase transition,''
Phys. Lett. B \textbf{780}, 313-318 (2018)
%doi:10.1016/j.physletb.2018.03.011
[arXiv:1609.07969 [hep-ph]].

\bibitem{Cohen:1990py}
A.~G.~Cohen, D.~B.~Kaplan and A.~E.~Nelson,
``WEAK SCALE BARYOGENESIS,''
Phys. Lett. B \textbf{245}, 561-564 (1990).

\bibitem{Cohen:1990it}
A.~G.~Cohen, D.~B.~Kaplan and A.~E.~Nelson,
``Baryogenesis at the weak phase transition,''
Nucl. Phys. B \textbf{349}, 727-742 (1991).

\bibitem{Rubakov:1997vp}
V.~A.~Rubakov,
``Grand unification and heavy axion,''
JETP Lett. \textbf{65}, 621-624 (1997)
%doi:10.1134/1.567390
[arXiv:hep-ph/9703409 [hep-ph]].

\bibitem{Berezhiani:2000gh}
Z.~Berezhiani, L.~Gianfagna and M.~Giannotti,
``Strong CP problem and mirror world: The Weinberg-Wilczek axion revisited,''
Phys. Lett. B \textbf{500}, 286-296 (2001)
%doi:10.1016/S0370-2693(00)01392-7
[arXiv:hep-ph/0009290 [hep-ph]].

\bibitem{Hook:2014cda}
A.~Hook,
``Anomalous solutions to the strong CP problem,''
Phys. Rev. Lett. \textbf{114}, no.14, 141801 (2015)
%doi:10.1103/PhysRevLett.114.141801
[arXiv:1411.3325 [hep-ph]].

\bibitem{Fukuda:2015ana}
H.~Fukuda, K.~Harigaya, M.~Ibe and T.~T.~Yanagida,
``Model of visible QCD axion,''
Phys. Rev. D \textbf{92}, no.1, 015021 (2015)
%doi:10.1103/PhysRevD.92.015021
[arXiv:1504.06084 [hep-ph]].

\bibitem{Crewther:1979pi}
R.~J.~Crewther, P.~Di Vecchia, G.~Veneziano and E.~Witten,
``Chiral Estimate of the Electric Dipole Moment of the Neutron in Quantum Chromodynamics,''
Phys. Lett. B \textbf{88}, 123 (1979)
[erratum: Phys. Lett. B \textbf{91}, 487 (1980)].

\bibitem{Baker:2006ts}
C.~A.~Baker, D.~D.~Doyle, P.~Geltenbort, K.~Green, M.~G.~D.~van der Grinten, P.~G.~Harris, P.~Iaydjiev, S.~N.~Ivanov, D.~J.~R.~May and J.~M.~Pendlebury, \textit{et al.}
``An Improved experimental limit on the electric dipole moment of the neutron,''
Phys. Rev. Lett. \textbf{97}, 131801 (2006)
%doi:10.1103/PhysRevLett.97.131801
[arXiv:hep-ex/0602020 [hep-ex]].

\bibitem{Kim:1979if}
J.~E.~Kim,
``Weak Interaction Singlet and Strong CP Invariance,''
Phys. Rev. Lett. \textbf{43}, 103 (1979).

\bibitem{Shifman:1979if}
M.~A.~Shifman, A.~I.~Vainshtein and V.~I.~Zakharov,
``Can Confinement Ensure Natural CP Invariance of Strong Interactions?,''
Nucl. Phys. B \textbf{166}, 493-506 (1980).

\bibitem{DiLuzio:2020wdo}
L.~Di Luzio, M.~Giannotti, E.~Nardi and L.~Visinelli,
``The landscape of QCD axion models,''
Phys. Rept. \textbf{870}, 1-117 (2020)
%doi:10.1016/j.physrep.2020.06.002
[arXiv:2003.01100 [hep-ph]].

\bibitem{VonHarling:2019rgb}
B.~Von Harling, A.~Pomarol, O.~Pujol\`as and F.~Rompineve,
``Peccei-Quinn Phase Transition at LIGO,''
JHEP \textbf{04}, 195 (2020)
%doi:10.1007/JHEP04(2020)195
[arXiv:1912.07587 [hep-ph]].

\bibitem{DelleRose:2019pgi}
L.~Delle Rose, G.~Panico, M.~Redi and A.~Tesi,
``Gravitational Waves from Supercool Axions,''
JHEP \textbf{04}, 025 (2020)
%doi:10.1007/JHEP04(2020)025
[arXiv:1912.06139 [hep-ph]].

\bibitem{Ghoshal:2020vud}
A.~Ghoshal and A.~Salvio,
``Gravitational waves from fundamental axion dynamics,''
JHEP \textbf{12}, 049 (2020)
%doi:10.1007/JHEP12(2020)049
[arXiv:2007.00005 [hep-ph]].
	
\bibitem{Coleman:1973jx}
S.~R.~Coleman and E.~J.~Weinberg,
``Radiative Corrections as the Origin of Spontaneous Symmetry Breaking,''
Phys. Rev. D \textbf{7}, 1888-1910 (1973).

\bibitem{Quiros:1999jp}
M.~Quiros,
``Finite temperature field theory and phase transitions,''
[arXiv:hep-ph/9901312 [hep-ph]].

\bibitem{Curtin:2016urg}
D.~Curtin, P.~Meade and H.~Ramani,
``Thermal Resummation and Phase Transitions,''
Eur. Phys. J. C \textbf{78}, no.9, 787 (2018)
%doi:10.1140/epjc/s10052-018-6268-0
[arXiv:1612.00466 [hep-ph]].

\bibitem{Masoumi:2016wot} 
A.~Masoumi, K.~D.~Olum and B.~Shlaer,
``Efficient numerical solution to vacuum decay with many fields,''
JCAP \textbf{01}, 051 (2017)
%doi:10.1088/1475-7516/2017/01/051
[arXiv:1610.06594 [gr-qc]].

\bibitem{Linde:1980tt} 
A.~D.~Linde,
``Fate of the False Vacuum at Finite Temperature: Theory and Applications,''
Phys. Lett. B \textbf{100}, 37-40 (1981).

\bibitem{Cline:2000nw}
J.~M.~Cline, M.~Joyce and K.~Kainulainen,
``Supersymmetric electroweak baryogenesis,''
JHEP \textbf{07}, 018 (2000)
%doi:10.1088/1126-6708/2000/07/018
[arXiv:hep-ph/0006119 [hep-ph]].

\bibitem{Fromme:2006wx}
L.~Fromme and S.~J.~Huber,
``Top transport in electroweak baryogenesis,''
JHEP \textbf{03}, 049 (2007)
%doi:10.1088/1126-6708/2007/03/049
[arXiv:hep-ph/0604159 [hep-ph]].

\bibitem{Bruggisser:2017lhc}
S.~Bruggisser, T.~Konstandin and G.~Servant,
``CP-violation for Electroweak Baryogenesis from Dynamical CKM Matrix,''
JCAP \textbf{11}, 034 (2017)
%doi:10.1088/1475-7516/2017/11/034
[arXiv:1706.08534 [hep-ph]].

\bibitem{Planck:2018vyg}
N.~Aghanim \textit{et al.} [Planck],
``Planck 2018 results. VI. Cosmological parameters,''
Astron. Astrophys. \textbf{641}, A6 (2020)
%doi:10.1051/0004-6361/201833910
[arXiv:1807.06209 [astro-ph.CO]].

\bibitem{Zeldovich:1974uw}
Y.~B.~Zeldovich, I.~Y.~Kobzarev and L.~B.~Okun,
``Cosmological Consequences of the Spontaneous Breakdown of Discrete Symmetry,''
Zh. Eksp. Teor. Fiz. \textbf{67}, 3-11 (1974)
SLAC-TRANS-0165.

\bibitem{Mazumdar:2018dfl}
A.~Mazumdar and G.~White,
``Review of cosmic phase transitions: their significance and experimental signatures,''
Rept. Prog. Phys. \textbf{82}, no.7, 076901 (2019)
%doi:10.1088/1361-6633/ab1f55
[arXiv:1811.01948 [hep-ph]].

\bibitem{Ahmadvand:2017xrw}
M.~Ahmadvand and K.~Bitaghsir Fadafan,
``Gravitational waves generated from the cosmological QCD phase transition within AdS/QCD,''
Phys. Lett. B \textbf{772}, 747-751 (2017)
%doi:10.1016/j.physletb.2017.07.039
[arXiv:1703.02801 [hep-th]].

\bibitem{Ahmadvand:2017tue}
M.~Ahmadvand and K.~Bitaghsir Fadafan,
``The cosmic QCD phase transition with dense matter and its gravitational waves from holography,''
Phys. Lett. B \textbf{779}, 1-8 (2018)
%doi:10.1016/j.physletb.2018.01.066
[arXiv:1707.05068 [hep-th]].

\bibitem{Abedi:2019msi}
H.~Abedi, M.~Ahmadvand and S.~S.~Gousheh,
``Electroweak phase transition in the presence of hypermagnetic field and the generation of gravitational waves,''
[arXiv:1901.05912 [hep-ph]].

\bibitem{Ahmadvand:2020fqv}
M.~Ahmadvand, K.~Bitaghsir Fadafan and S.~Rezapour,
``Gravitational waves of a first-order QCD phase transition at finite coupling from holography,''
[arXiv:2006.04265 [hep-th]].

		\bibitem{Kosowsky:1992rz}
A.~Kosowsky, M.~S.~Turner and R.~Watkins,
``Gravitational waves from first order cosmological phase transitions,''
Phys. Rev. Lett. \textbf{69}, 2026-2029 (1992).

\bibitem{Kamionkowski:1993fg} 
M.~Kamionkowski, A.~Kosowsky and M.~S.~Turner, 
``Gravitational radiation from first order phase transitions,''
Phys.\ Rev.\ D {\bf 49}, 2837 (1994) 
[astro-ph/9310044].

\bibitem{Kosowsky:2001xp}
A.~Kosowsky, A.~Mack and T.~Kahniashvili,
``Gravitational radiation from cosmological turbulence,''
Phys. Rev. D \textbf{66}, 024030 (2002)
[arXiv:astro-ph/0111483 [astro-ph]].

\bibitem{Hindmarsh:2015qta} 
M.~Hindmarsh, S.~J.~Huber, K.~Rummukainen and D.~J.~Weir,
``Numerical simulations of acoustically generated gravitational waves at a first order phase transition,'' 
Phys.\ Rev.\ D {\bf 92}, no. 12, 123009 (2015) 
[arXiv:1504.03291 [astro-ph.CO]].

\bibitem{Caprini:2015zlo}
C.~Caprini, M.~Hindmarsh, S.~Huber, T.~Konstandin, J.~Kozaczuk, G.~Nardini, J.~M.~No, A.~Petiteau, P.~Schwaller and G.~Servant, \textit{et al.}
``Science with the space-based interferometer eLISA. II: Gravitational waves from cosmological phase transitions,''
JCAP \textbf{04}, 001 (2016)
%doi:10.1088/1475-7516/2016/04/001
[arXiv:1512.06239 [astro-ph.CO]].

\bibitem{Caprini:2009yp}
C.~Caprini, R.~Durrer and G.~Servant,
``The stochastic gravitational wave background from turbulence and magnetic fields generated by a first-order phase transition,''
JCAP \textbf{12}, 024 (2009)
%doi:10.1088/1475-7516/2009/12/024
[arXiv:0909.0622 [astro-ph.CO]].

\bibitem{Espinosa:2010hh}
J.~R.~Espinosa, T.~Konstandin, J.~M.~No and G.~Servant,
``Energy Budget of Cosmological First-order Phase Transitions,''
JCAP \textbf{06}, 028 (2010)
%doi:10.1088/1475-7516/2010/06/028
[arXiv:1004.4187 [hep-ph]].

\bibitem{Hild:2008ng}
S.~Hild, S.~Chelkowski and A.~Freise,
``Pushing towards the ET sensitivity using 'conventional' technology,''
[arXiv:0810.0604 [gr-qc]].

\bibitem{Sathyaprakash:2012jk}
B.~Sathyaprakash, M.~Abernathy, F.~Acernese, P.~Ajith, B.~Allen, P.~Amaro-Seoane, N.~Andersson, S.~Aoudia, K.~Arun and P.~Astone, \textit{et al.}
``Scientific Objectives of Einstein Telescope,''
Class. Quant. Grav. \textbf{29}, 124013 (2012)
[erratum: Class. Quant. Grav. \textbf{30}, 079501 (2013)]
%doi:10.1088/0264-9381/29/12/124013
[arXiv:1206.0331 [gr-qc]].

\bibitem{DiLuzio:2021pxd}
L.~Di Luzio, B.~Gavela, P.~Quilez and A.~Ringwald,
``An even lighter QCD axion,''
JHEP \textbf{05}, 184 (2021)
%doi:10.1007/JHEP05(2021)184
[arXiv:2102.00012 [hep-ph]].

\bibitem{Kudoh:2005as}
H.~Kudoh, A.~Taruya, T.~Hiramatsu and Y.~Himemoto,
``Detecting a gravitational-wave background with next-generation space interferometers,''
Phys. Rev. D \textbf{73}, 064006 (2006)
%doi:10.1103/PhysRevD.73.064006
[arXiv:gr-qc/0511145 [gr-qc]].

\bibitem{Yagi:2011wg}
K.~Yagi and N.~Seto,
``Detector configuration of DECIGO/BBO and identification of cosmological neutron-star binaries,''
Phys. Rev. D \textbf{83}, 044011 (2011)
[erratum: Phys. Rev. D \textbf{95}, no.10, 109901 (2017)]
%doi:10.1103/PhysRevD.83.044011
[arXiv:1101.3940 [astro-ph.CO]].

\bibitem{Smith:2006ym}
M.~C.~Smith, G.~R.~Ruchti, A.~Helmi, R.~F.~G.~Wyse, J.~P.~Fulbright, K.~C.~Freeman, J.~F.~Navarro, G.~M.~Seabroke, M.~Steinmetz and M.~Williams, \textit{et al.}
``The RAVE Survey: Constraining the Local Galactic Escape Speed,''
Mon. Not. Roy. Astron. Soc. \textbf{379}, 755-772 (2007)
%doi:10.1111/j.1365-2966.2007.11964.x
[arXiv:astro-ph/0611671 [astro-ph]].

\bibitem{Kannike:2020agf}
K.~Kannike, M.~Raidal, H.~Veerm\"ae, A.~Strumia and D.~Teresi,
``Dark Matter and the XENON1T electron recoil excess,''
Phys. Rev. D \textbf{102}, no.9, 095002 (2020)
%doi:10.1103/PhysRevD.102.095002
[arXiv:2006.10735 [hep-ph]].

\bibitem{XENON:2020rca}
E.~Aprile \textit{et al.} [XENON],
``Excess electronic recoil events in XENON1T,''
Phys. Rev. D \textbf{102}, no.7, 072004 (2020)
%doi:10.1103/PhysRevD.102.072004
[arXiv:2006.09721 [hep-ex]].

\bibitem{Shakeri:2020wvk}
S.~Shakeri, F.~Hajkarim and S.~S.~Xue,
``Shedding New Light on Sterile Neutrinos from XENON1T Experiment,''
JHEP \textbf{12}, 194 (2020)
%doi:10.1007/JHEP12(2020)194
[arXiv:2008.05029 [hep-ph]].

\bibitem{Arguelles:2019ouk}
C.~A.~Arg\"uelles, A.~Diaz, A.~Kheirandish, A.~Olivares-Del-Campo, I.~Safa and A.~C.~Vincent,
``Dark Matter Annihilation to Neutrinos,''
[arXiv:1912.09486 [hep-ph]].

\bibitem{Chianese:2021htv}
M.~Chianese, D.~F.~G.~Fiorillo, R.~Hajjar, G.~Miele, S.~Morisi and N.~Saviano,
``Heavy decaying dark matter at future neutrino radio telescopes,''
JCAP \textbf{05}, 074 (2021)
%doi:10.1088/1475-7516/2021/05/074
[arXiv:2103.03254 [hep-ph]].

% Please avoid comments such as "For a review'', "For some examples",
% "and references therein" or move them in the text. In general,
% please leave only references in the bibliography and move all
% accessory text in footnotes.

% Also, please have only one work for each \bibitem.


\end{thebibliography}
\end{document}